\documentclass[conference]{IEEEtran}
\IEEEoverridecommandlockouts
\usepackage{cite}
\usepackage{amsmath,amssymb,amsfonts}
\usepackage{algorithmic}
\usepackage{graphicx}
\usepackage{textcomp}
\usepackage{xcolor}
\usepackage{subcaption}
\def\BibTeX{{\rm B\kern-.05em{\sc i\kern-.025em b}\kern-.08em
    T\kern-.1667em\lower.7ex\hbox{E}\kern-.125emX}}
\begin{document}

\title{Novel Approach to Dual-Channel Estimation in Integrated Sensing and Communications for 6G\\

}

\author{\IEEEauthorblockN{Alejandro Castilla; Saúl Fenollosa; Monika Drozdowska; \\
Alejandro Lopez-Escudero; Sergio Micò-Rosa;
Narcis Cardona}
\IEEEauthorblockA{\textit{iTEAM Research Institute; Universitat Politècnica de València, Valencia, Spain} \\
\textit{[a.castilla, sjfenarg, mdrozdo, alloes3, sermiro, ncardona]@upv.edu.es}}
}

\maketitle

\begin{abstract}
Integrated Sensing and Communication (ISAC) design is crucial for 6G and harmonizes environmental data sensing with communication, emphasizing the need to understand and model these elements. This paper delves into dual-channel models for ISAC, employing channel extraction techniques to validate and enhance accuracy. Focusing on millimeter wave (mmWave) radars, it explores the extraction of the bistatic sensing channel from monostatic measurements and subsequent communication channel estimation. The proposed methods involve interference extraction, module and phase correlation analyses, chirp clustering, and auto-clutter reduction. A comprehensive set-up in an anechoic chamber with controlled scenarios evaluates the proposed techniques, demonstrating successful channel extraction and validation through Root Mean Square Delay Spread (RMS DS), Power Delay Profile (PDP), and Angle of Arrival (AoA) analysis. Comparison with Ray-Tracing (RT) simulations confirms the effectiveness of the proposed approach, presenting an innovative stride towards fully integrated sensing and communication in future networks.
\end{abstract}


\section{Introduction}
The upcoming Next Generation Communication Networks (6G) are expected to integrate communication, sensing, and localization techniques at exceptionally high frequencies \cite{b1}. As the channel coherence time is expected to decrease and radio variations occur more frequently, new dynamic radio channel models will be necessary to capture the channel impulse response accurately. This highlights the need for innovative and effective methods for estimating radio channel properties \cite{b2}. Combining sensing and communication capabilities within a unified wireless framework allows environmental data to be sensed and collected alongside communication capabilities, enabling dynamic adjustments and performance optimization based on the collected information. This ISAC design paradigm has become crucial for 6G, emphasizing the importance of understanding and modeling the behavior of sensing and communication elements. 

The integration of sensing and communication systems is becoming increasingly prevalent in specific 5G verticals, such as vehicular communication systems \cite{b5,b6,b7}. Several studies have been conducted to explore ways of utilizing sensing channel information to improve communication channels. In \cite{b10}, communication receivers were studied, acting as targets in a monostatic orthogonal time-frequency space-based (OTFS) ISAC system. The authors successfully predicted delay and Doppler parameters by leveraging similarities between the communication and sensing channels. In \cite{b11}, OFTS modulation was employed to estimate radar parameters and communication signals jointly. This method leverages channel sparsity in the Delay-Doppler domain, enabling efficient separation of radar parameter estimation and communication signals. However, implementing monostatic ISAC sensing requires the full-duplex mode of the radio interface, which is currently not incorporated into the current 3GPP standards. Therefore, many proposed ISAC systems have adopted a JSAC (Joint Sensing and Communications) strategy, which uses a radar sensor to enhance communication links.

Future networks benefit from fully integrating sensing and communications in a unified radio signal for enhanced spectral efficiency \cite{b11}. This ISAC approach utilizes a single modulation scheme for concurrent communication data and environmental perception. However, it's crucial to recognize that the behavior of sensing and communication channels, despite sharing the same frequency spectrum, can't be universally modeled similarly. Varied strategies are pursued, including new propagation models based on environmental intelligence \cite{b12} and creating digital twins for synthetically simulated channel data. While stochastic point cloud models in 3GPP model communication channel scattering \cite{b13}, incorporating the sensing channel into 3GPP standards remains unaddressed, making dual-channel ISAC modeling a hot topic for research.

This paper aims to contribute to defining dual channel models for ISAC and how they relate. To achieve this goal, we employ the channel extraction techniques outlined in the research papers referenced as \cite{b14} and \cite{b15}. By utilizing radars in the mmWave band, the papers demonstrate the extraction of the bistatic sensing channel from a monostatic measurement, enabling the estimation of the communications channel. This is possible due to the similarity of the communications and bistatic sensing channels when using the same frequency band. The present work aims to validate and examine the accuracy of these channel extraction methods and describe the recent enhancements to the referenced algorithms.

\section{ESTIMATION OF COMMUNICATIONS CHANNEL FROM MONOSTATIC SENSING MEASUREMENTS AT mmWAVE BAND}\label{algorithm}

\begin{figure} 
\centering
\begin{subfigure}{.5\textwidth}
  \centering
  \includegraphics[width=.9\linewidth]{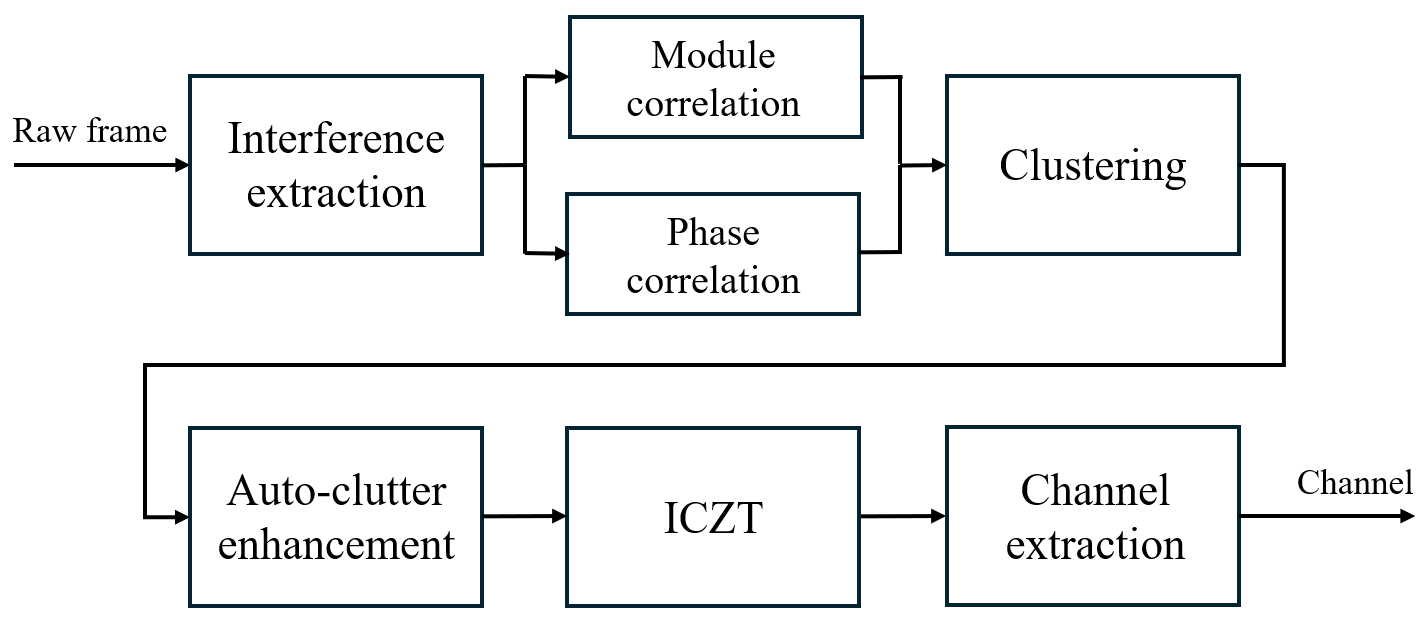}
\end{subfigure}%
\caption{Block diagram of the channel extraction processing chain.}
\label{fig:block_diagram}
\end{figure}

Based on the research presented in \cite{b14,b15}, it has been found that mutual interference between two monostatic frequency-modulated continuous wave (FMCW) radars can contain valuable information that can be used to extract the communication channel. In \cite{b14}, the method of generating FMCW frequency ramps with unequal chirp slopes is described. This difference in slope generates a "ghost target" interference signal at each victim radar receiver, which can be used to estimate the communications channel. This section proposes significant improvements regarding the signal processing of \cite{b15}, where simplistic chirp pulse aggregation is performed. However, the time between samples is higher than the time a wavelength takes to do a cycle, causing every chirp to be sampled with different time references, impeding the direct coherent aggregation of chirp pulses. It is also essential to note that interference appears in multiple chirps at the receiver of the victim radar. Therefore, utilizing sampling diversity is crucial to obtain complete channel information. Figure \ref{fig:block_diagram} outlines extracting channels from a raw radar frame, and every subsection corresponds with each block in the figure.

As the radars can measure many chirps quickly, and most contain interference, statistically, some chirps have been sampled with similar phase references, allowing them to be aggregated. To this end, a dual analysis is applied to check the coherence between chirps. This is performed in subsections \ref{mod_corr} and \ref{ang_corr}. Once we correlate chirps, a clustering technique is applied in section \ref{clustering} to identify groups of coherent chirps. Each group contains part of the subsampled signal, to compose all the uncoherent chirps into one signal, an auto-clutter technique is applied in subsection \ref{Auto-clutter}. Finally, inverse Chirp Z-Transform (ICZT) is performed in subsection \ref{IZCT} to improve the resolution, then the channel can be extracted by averaging as explained in subsection \ref{Channel extraction}.
\subsection{Interference extraction}
\begin{figure}
\centering
\begin{subfigure}{.25\textwidth}
  \centering
  \includegraphics[width=.9\linewidth]{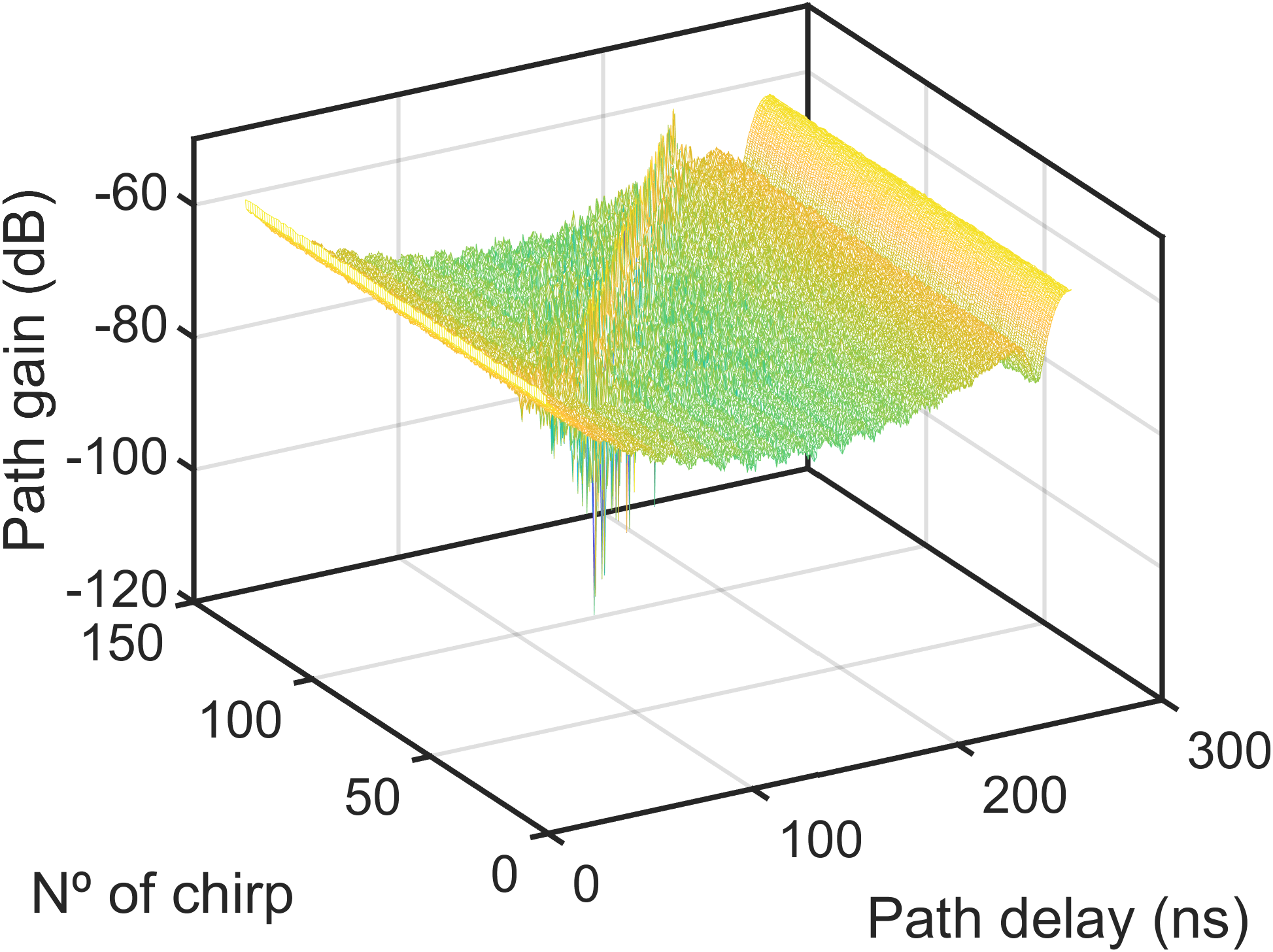}
  \caption{Raw frame.}
  \label{fig:raw_frame}
\end{subfigure}%
\begin{subfigure}{.25\textwidth}
  \centering
  \includegraphics[width=.9\linewidth]{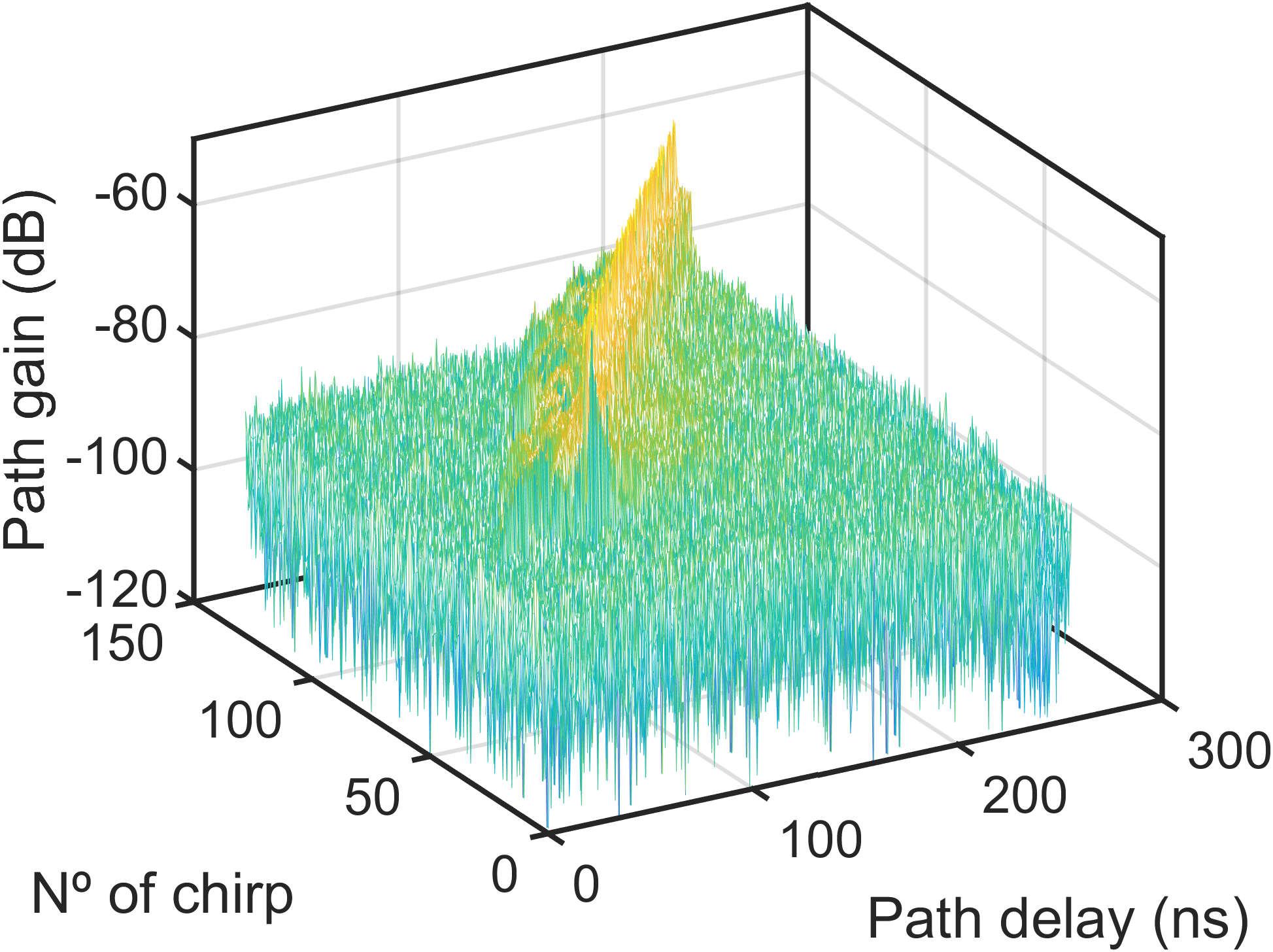}
  \caption{Static filtered frame.}
  \label{fig:filtered_frame}
\end{subfigure}
\vspace{2pt}
\newline
\begin{subfigure}{.23\textwidth}
  \centering
  \includegraphics[width=.9\linewidth]{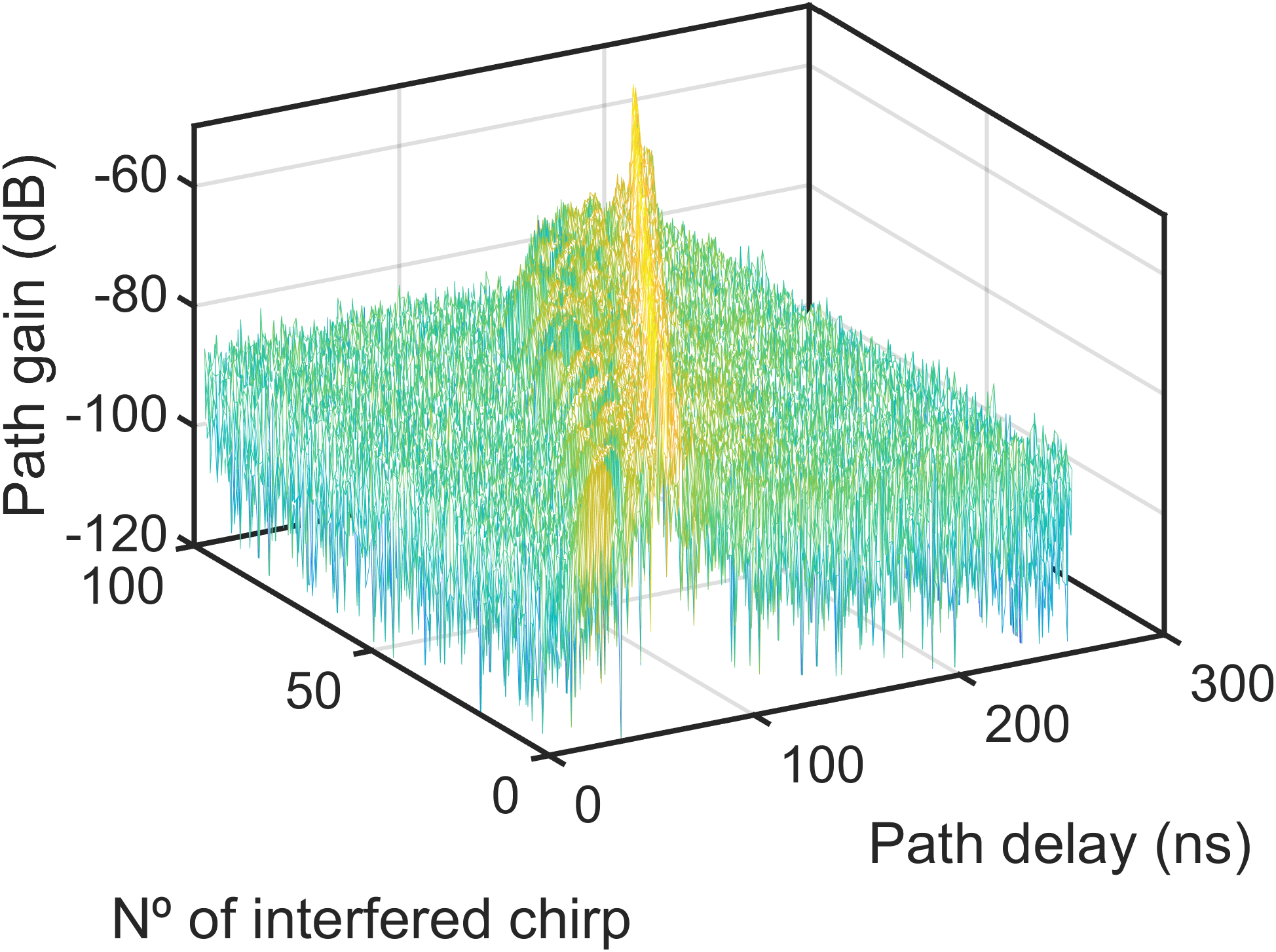}
  \caption{Interference chirps selection.}
  \label{fig:interference}
\end{subfigure}
\begin{subfigure}{.23\textwidth}
  \centering
  \includegraphics[width=.9\linewidth]{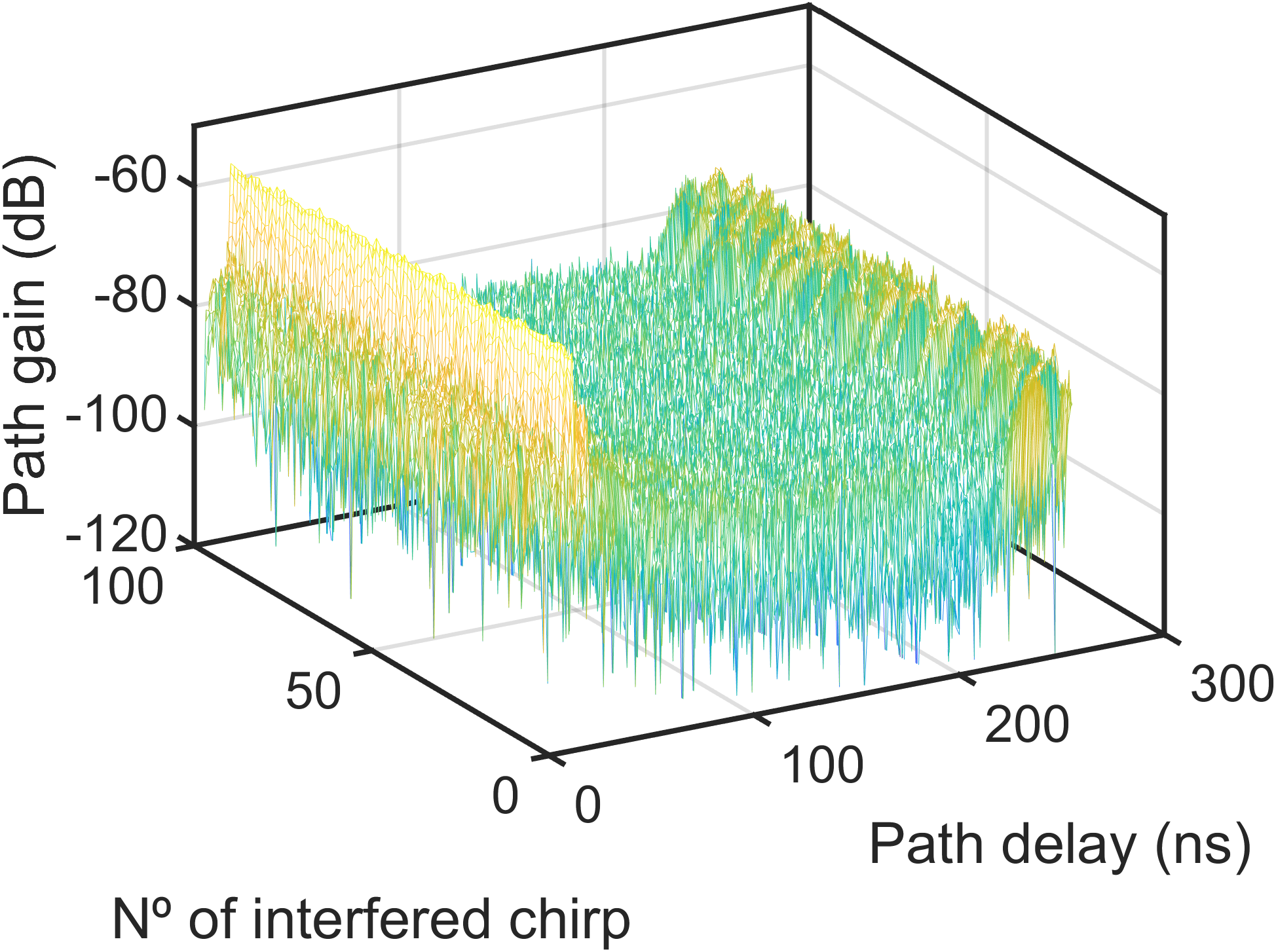}
  \caption{Chirp alignment.}
  \label{fig:interference_aligned}
\end{subfigure}
\caption{PDP of every phase in the interference extraction process.}
\label{fig:frame_signal_processing}
\end{figure}

The process of estimating channels from radar sensor's mutual interference starts with the extraction of the interference effect. This involves filtering out the static components to isolate the interference as performed in \cite{b15}. Chirps containing interference are extracted by applying a power threshold, and then preliminary alignment is performed, shifting every chirp to match the main contribution with the reference distance. Figure \ref{fig:frame_signal_processing} describes the entire extraction process.

\subsection{Module correlation}\label{mod_corr}

The following step involves examining the correlation between the modules of each chirp with the rest of them in the interference. If the correlation is sufficiently high, it indicates that those chirps have sampled similar channel information. The correlation values typically range from 0.7 to 0.98, and the highest correlations are utilized to identify the chirps that were sampled identically. Figure \ref{fig:mod_cor} provides an example of this outcome. 

\subsection{Phase correlation}\label{ang_corr}

\begin{figure}
\centering
\begin{subfigure}{.24\textwidth}
  \centering
  \includegraphics[width=\linewidth]{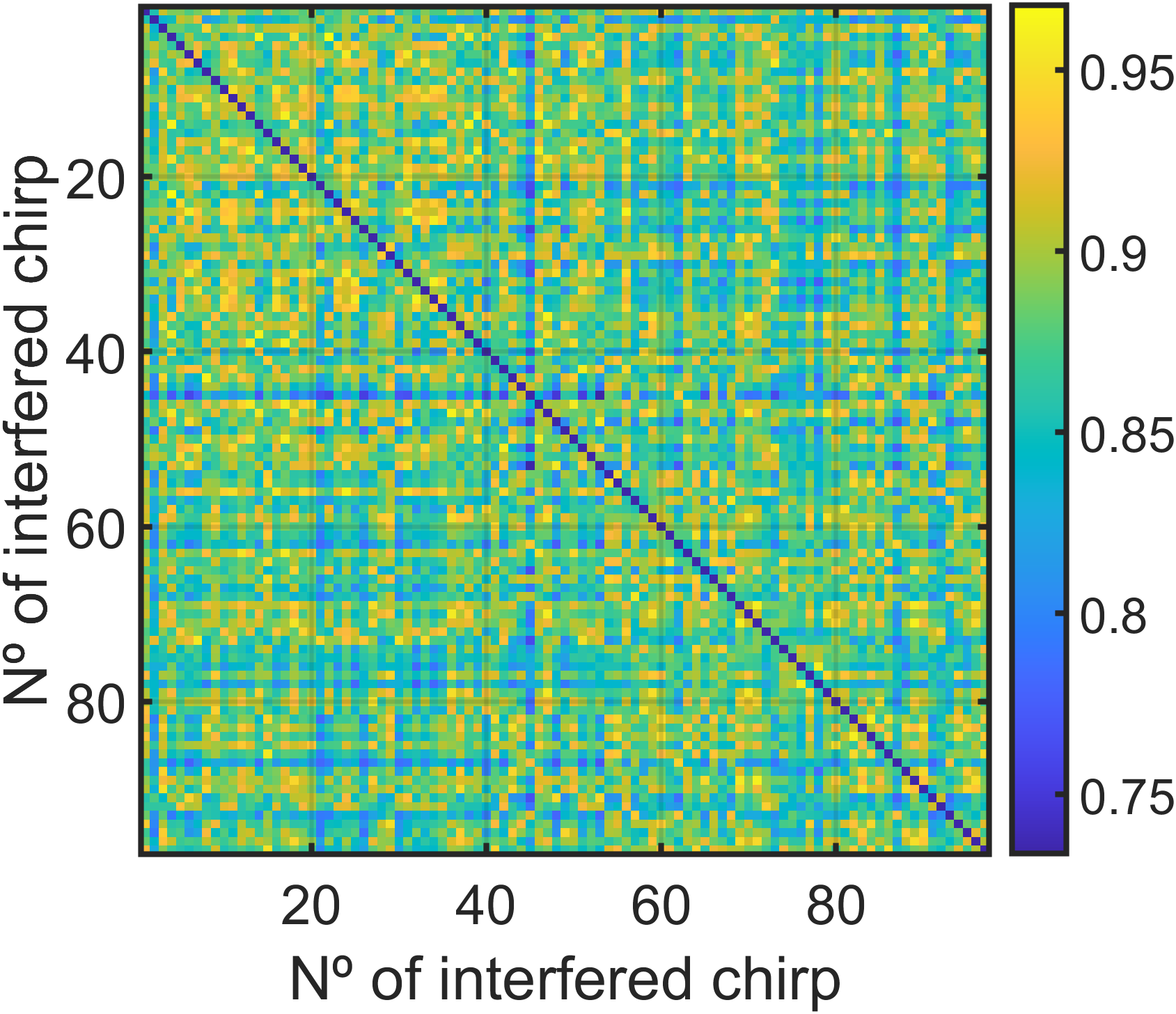}
  \caption{Module correlation.}
  \label{fig:mod_cor}
\end{subfigure}%
\hspace{1pt}
\begin{subfigure}{.233\textwidth}
  \centering
  \includegraphics[width=\linewidth]{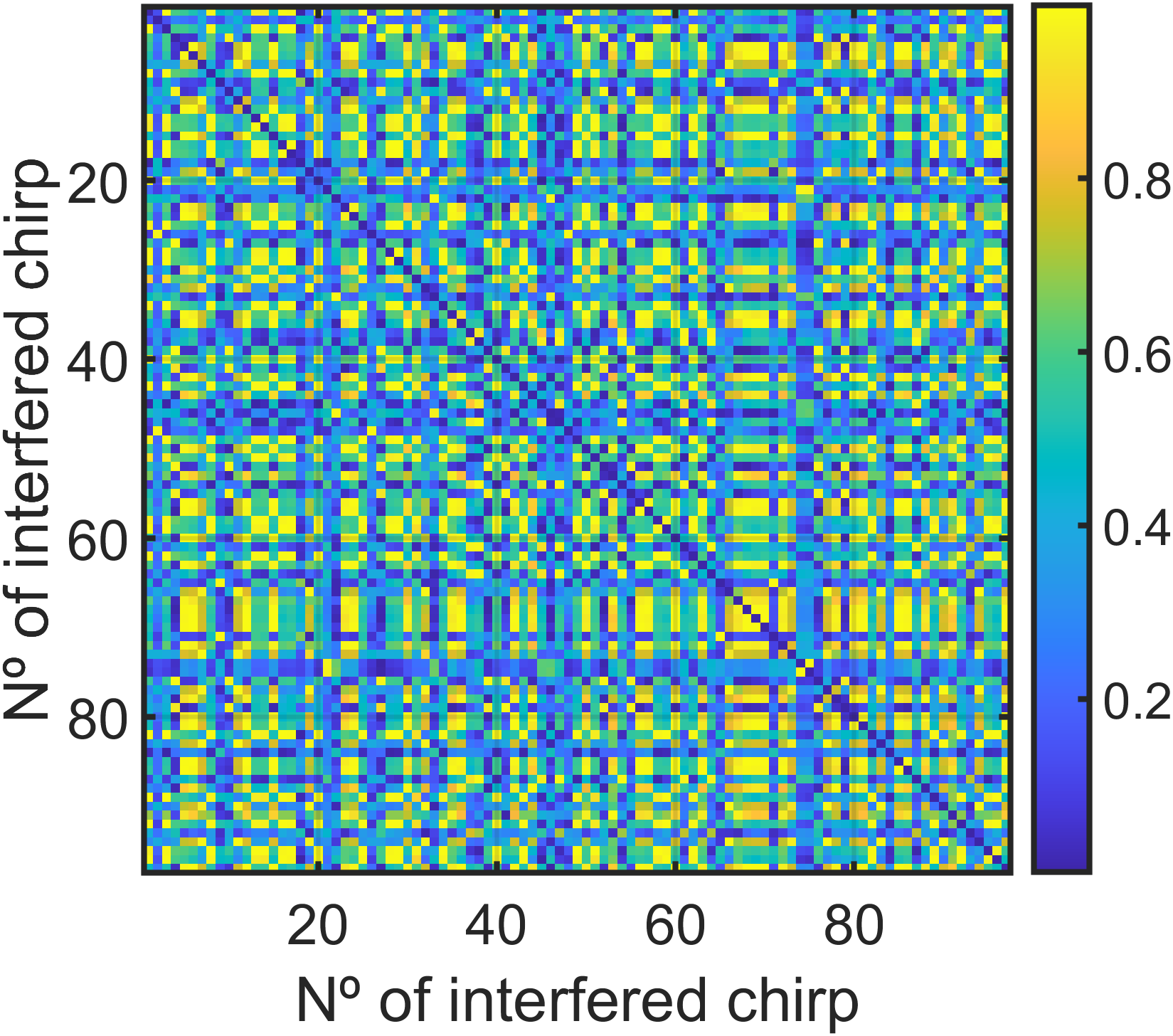}
  \caption{Phase correlation.}
  \label{fig:ang_cor}
\end{subfigure}
\caption{Correlation analysis between chirps.}
\label{fig:cor_analysis}
\end{figure}

Aside from module analysis, it is also essential to evaluate the phase consistency of various chirps. Phase variation can result in the loss of information due to destructive waves. To compare the phases of different chirps, the most powerful phase component in power must be obtained to ensure that it is part of the channel. Phase-coherent chirps can be identified by conducting a line-of-sight (LOS) phase chirp-to-chirp correlation. A representative case is presented in Figure \ref{fig:ang_cor}.

\subsection{Chirp clustering}\label{clustering}

\begin{figure}
\centering
\begin{subfigure}{.5\textwidth}
  \centering
  \includegraphics[width=.9\linewidth]{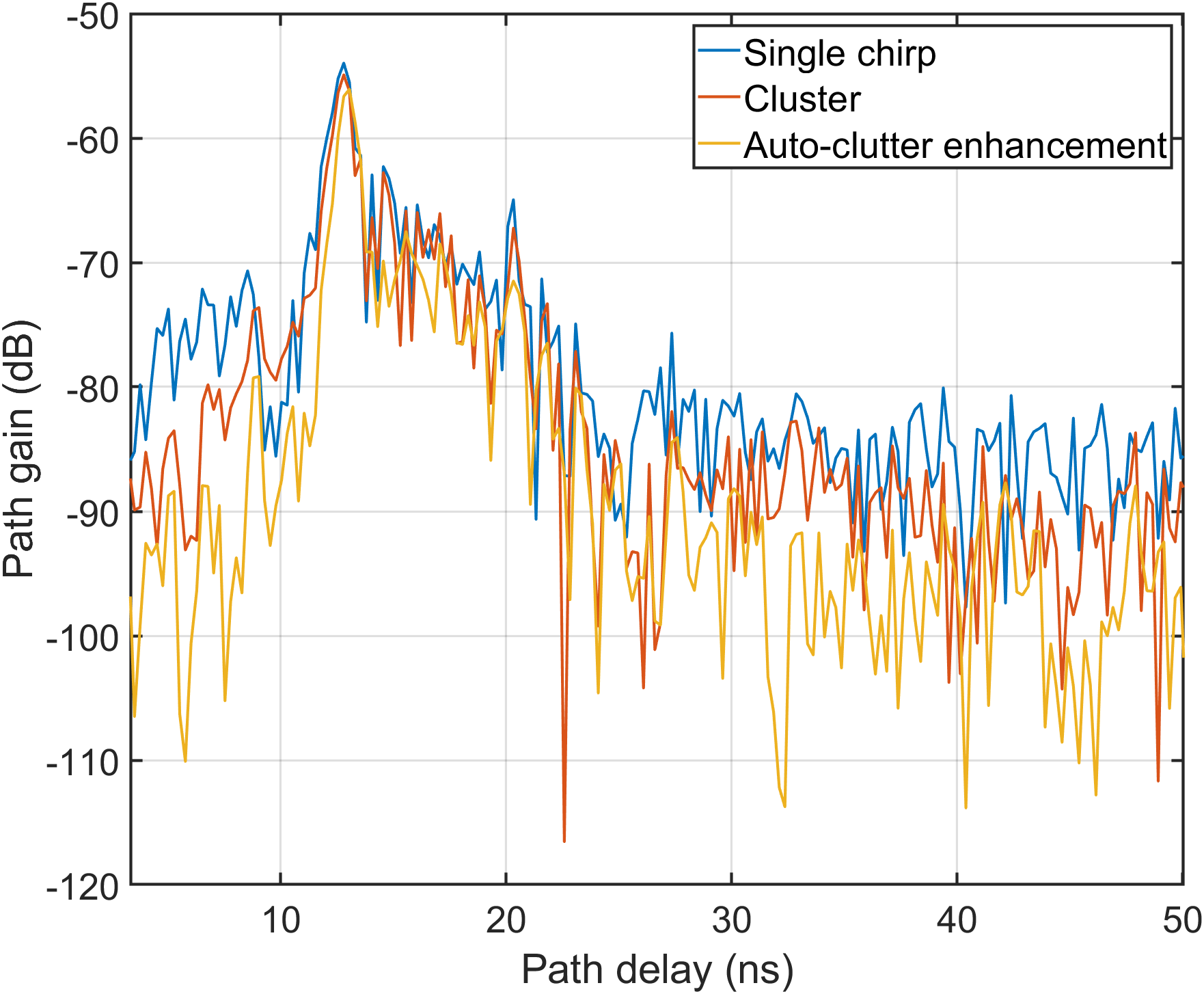}
\end{subfigure}%
\caption{Interfered chirps processing.}
\label{fig:chirps_processing}
\end{figure}

Coherent chirps are those with a high correlation in module and phase. These chirps are clustered by averaging their complex values. The result after clustering significantly improves the signal-to-noise (SNR) and allows the distinguishing of channel contributions from noise. The improvement after clustering can be seen in Figure \ref{fig:chirps_processing}.

\subsection{Auto-clutter reduction}\label{Auto-clutter}

Due to the clustering operation, each cluster must possess distinct channel information. However, the primary contribution is often of a magnitude that can give rise to secondary lobes through digital processing, which can mask other channel contributions. There are various auto-clutter enhancement techniques available, and in this particular scenario, we will employ the Hamming window in the frequency domain to mitigate secondary lobes. The outcome of this process must be compensated for by 5.35 dB to ensure the reliability of power information. The auto-clutter enhancement technique enables us to differentiate new components in the channel, and by diminishing the secondary lobes, the power values obtained for each component will be more reliable.

\subsection{Inverse Chirp Z-Transform}\label{IZCT}

\begin{figure}
\centering
\begin{subfigure}{.5\textwidth}
  \centering
  \includegraphics[width=.9\linewidth]{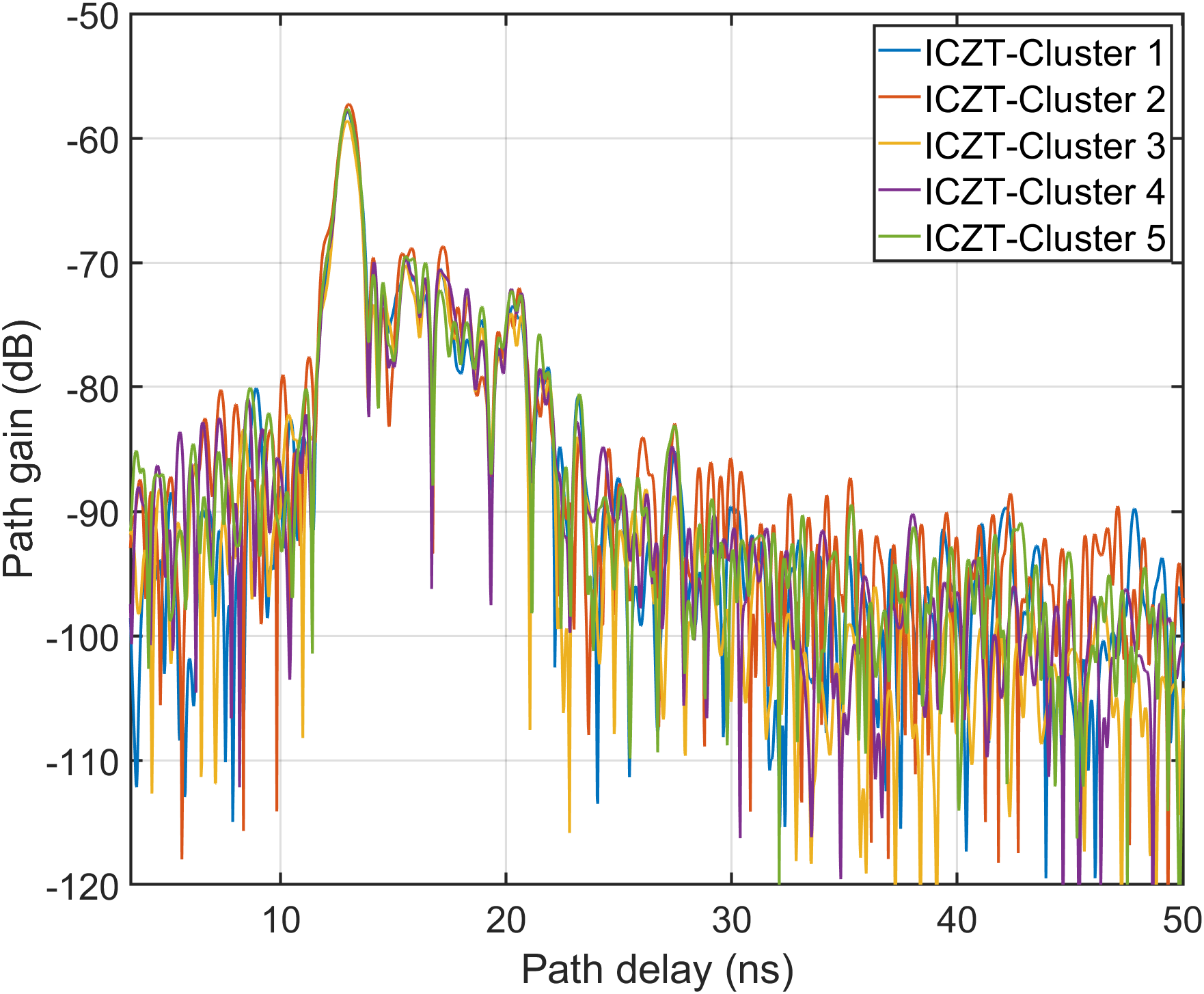}
\end{subfigure}%
\caption{ICZT obtained clusters.}
\label{fig:ICZT_clustering}
\end{figure}

Up to this point, the endeavour has been to gather as much information as possible without resorting to generalizations. However, the available data comprises numerous clusters that exhibit inconsistency with each other, albeit each cluster contains channel information. Precise alignment in the module and phase is vital to fully utilizing this information. By applying the ICZT in the frequency domain of each cluster, the channel component can be distinguished clearly thanks to the interpolation applied. The phase differences are also considered to prevent any loss of phase information. Once the range resolution is enhanced, the clusters are realigned by amplitude and phase to achieve coherence between clusters. The outcome of this ICZT-improved and re-aligned clustering is depicted in Figure \ref{fig:ICZT_clustering}.

\subsection{Channel extraction}\label{Channel extraction}
Finally, the channel impulse response (CIR) can be obtained by taking the average of the complex values for each ICZT-improved cluster.

\section{MEASUREMENT CAMPAIGN SET-UP AND GEOMETRICAL MULTIPATH DESCRIPTION}
\subsection{RF Hardware description} \label{RF_descr}

\begin{figure}
\centering
\begin{subfigure}{.5\textwidth}
  \centering
  \includegraphics[width=.9\linewidth]{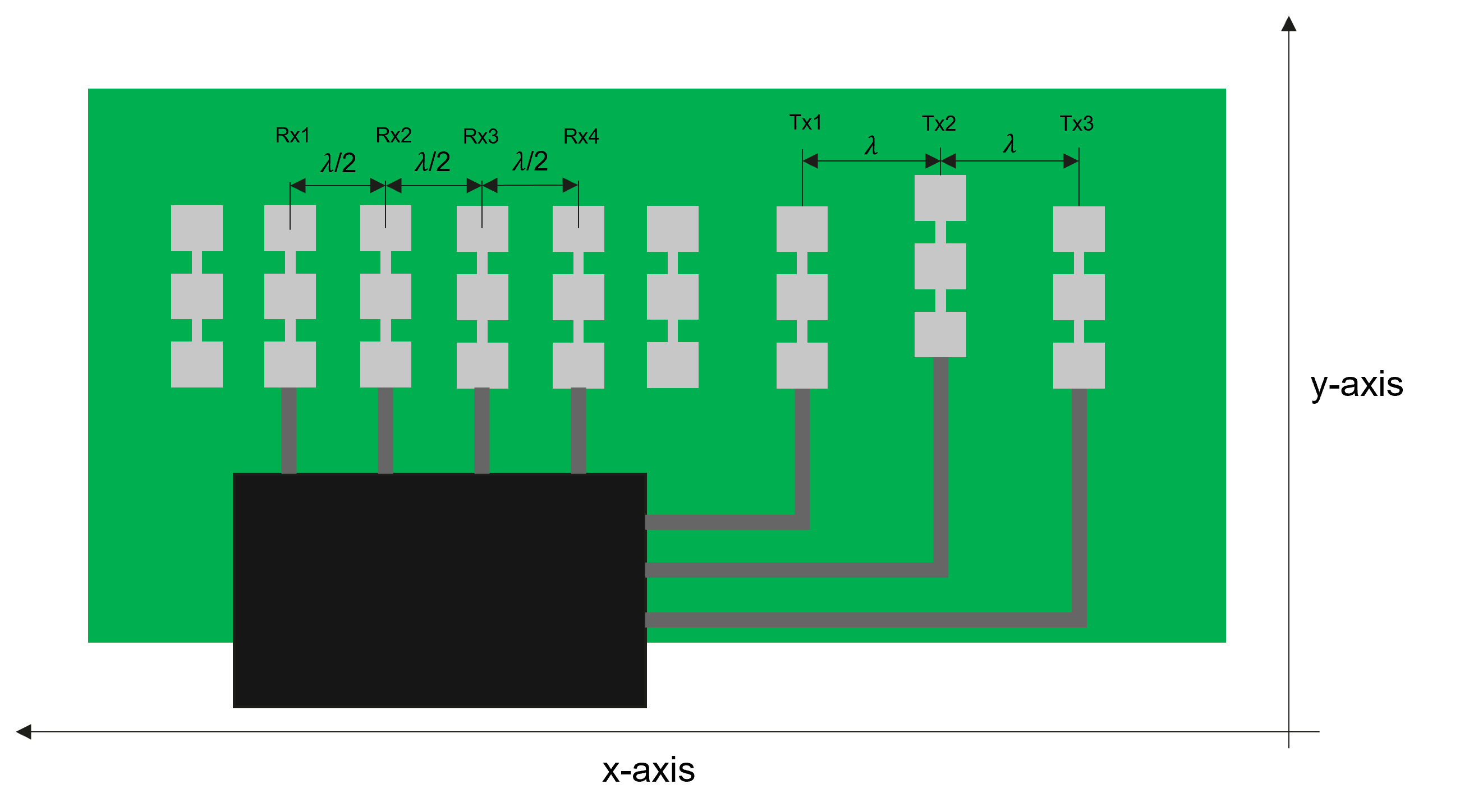}
\end{subfigure}%
\caption{mmWave radar antennas.}
\label{fig:radarAntennas}
\end{figure}

The radar employed in this paper operates within the mmWave frequency band spanning from 77 to 81 GHz, with a transmitter output power of up to 12 dBm and a receiver noise figure typically at 12 dB. The radar antenna comprises an array of four antennas, each with a half-wavelength spacing. Each antenna consists of three microstrip patches and is processed independently by a line, allowing for beamforming in the plane perpendicular to the patches. To ensure that all antennas are under the same conditions and that there are no boundary issues, two parasite antennas are connected to the ground on both sides of antennas 1 and 4. The radar has three transmitters spaced one wavelength apart and is shaped by a 3-patch microstrip line like the receivers. Referring to Figure \ref{fig:radarAntennas}, the x-axis has a beamwidth of 120$^\circ$ and is suitable for beamforming. In contrast, the y-axis has a beamwidth of 24$^\circ$ and does not allow for the application of beamforming algorithms. To determine the scenario's angle of arrival (AoA), the radar was positioned to enable beamforming with the ground during measurements. Hence, the AoA corresponded to the elevation angle of the received multipath. The elevation AoA will be used to verify the correct extraction of the multipath components' geometry of the communication channel.

\subsection{Description of geometrical multipath set-up configuration}\label{set-up_desc}

\begin{figure}
\centering
\begin{subfigure}{.21\textwidth}
  \centering
  \includegraphics[width=\linewidth]{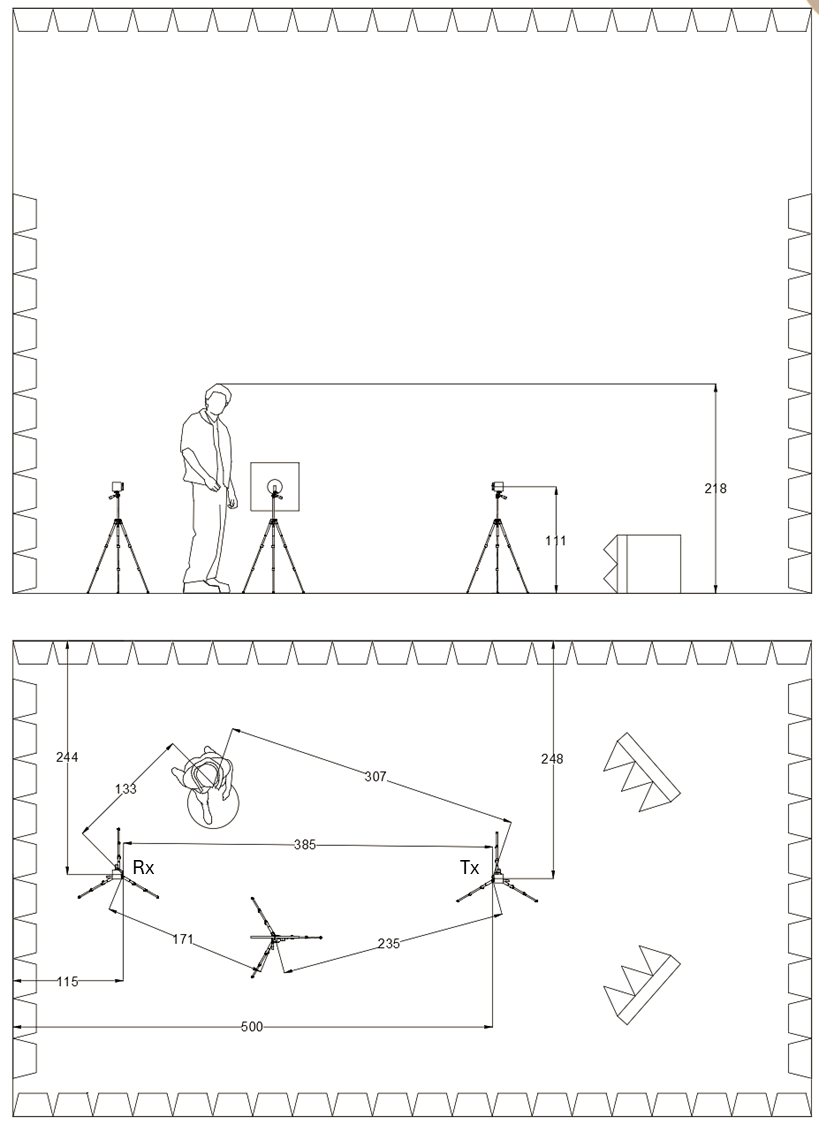}
  \caption{Schematic.}
  \label{fig:sh_S1_S2}
\end{subfigure}%
\begin{subfigure}{.267\textwidth}
  \centering
  \includegraphics[width=\linewidth]{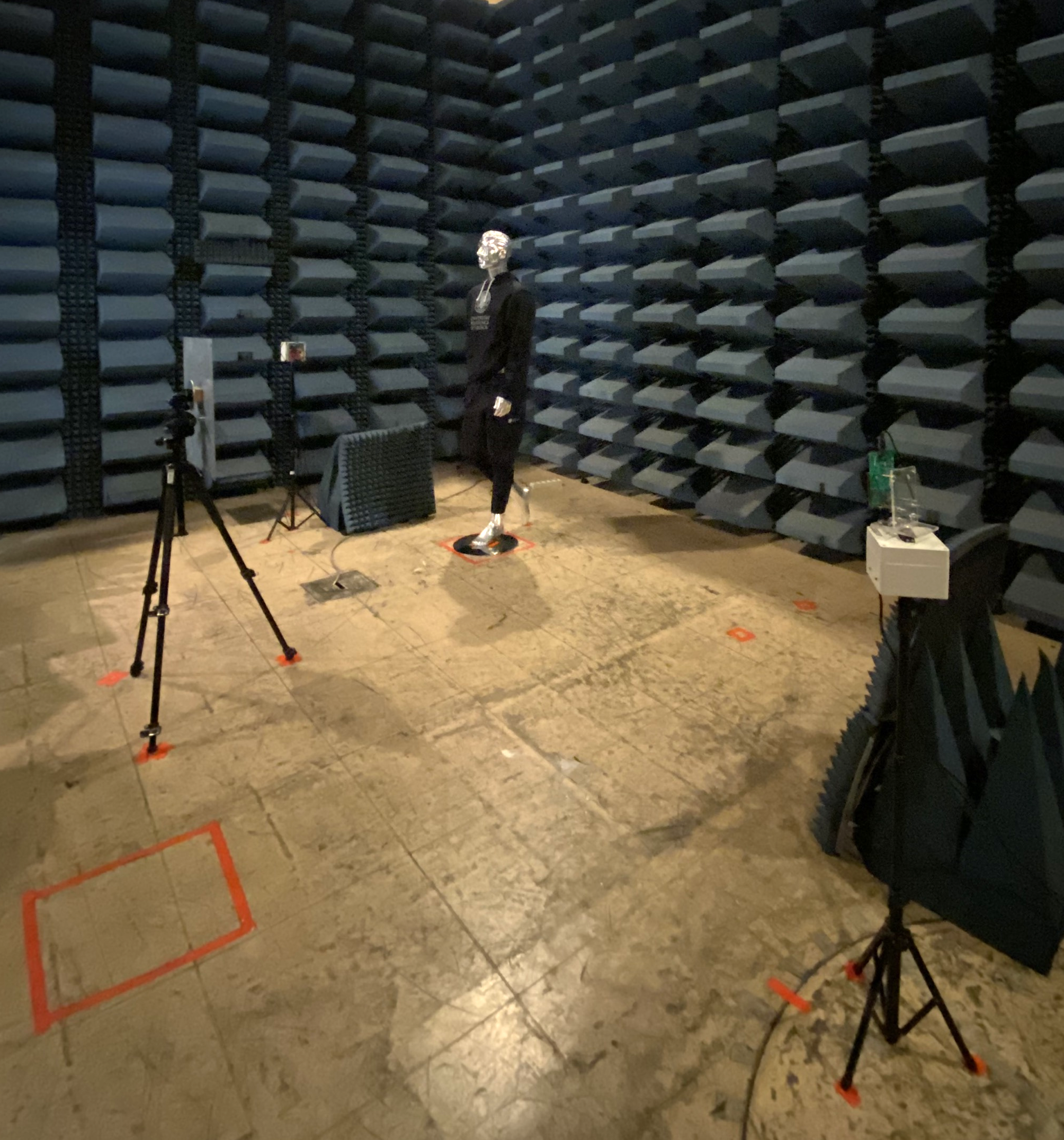}
  \caption{Set-up photography.}
  \label{fig:ph_S1_S2}
\end{subfigure}
\caption{Measurement set-up 1 (Set-up 2 is identical but without the metal plate scatterer).}
\label{fig:S1_S2}
\end{figure}
\begin{figure}
\centering
\begin{subfigure}{.21\textwidth}
  \centering
  \includegraphics[width=\linewidth]{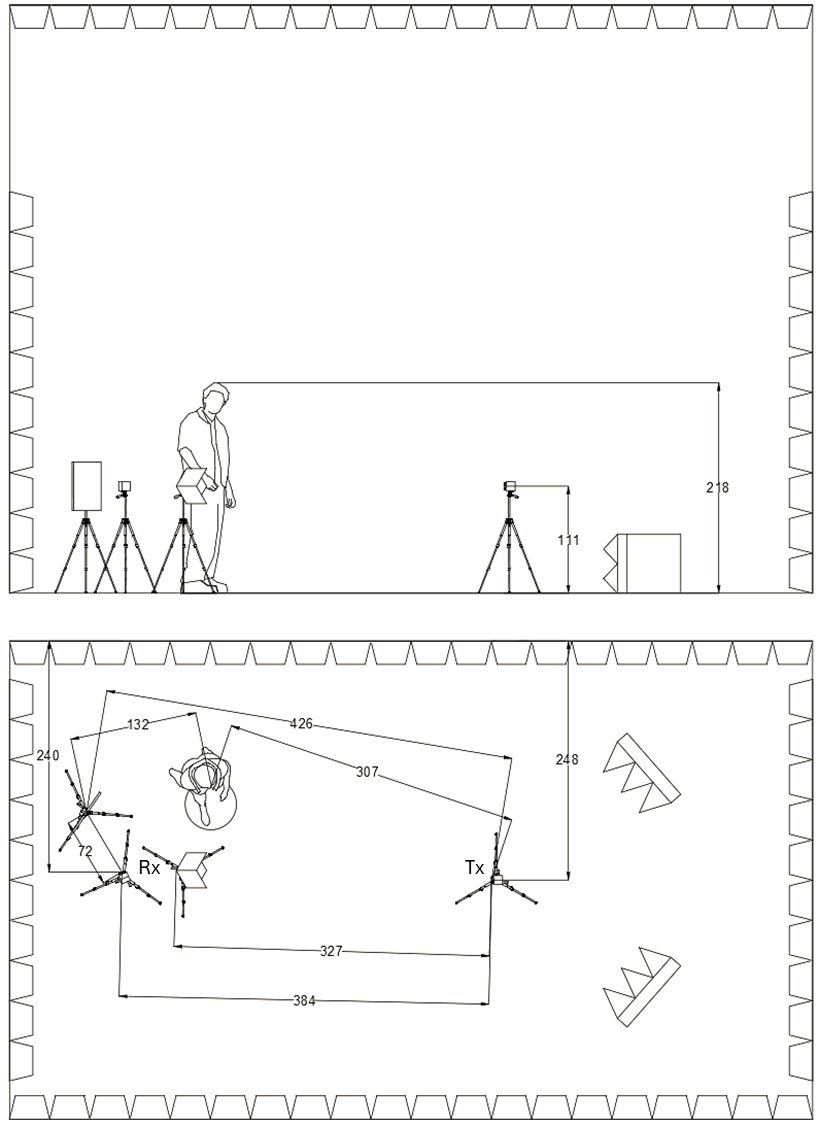}
  \caption{Schematic.}
  \label{fig:sh_S3}
\end{subfigure}%
\begin{subfigure}{.27\textwidth}
  \centering
  \includegraphics[width=\linewidth]{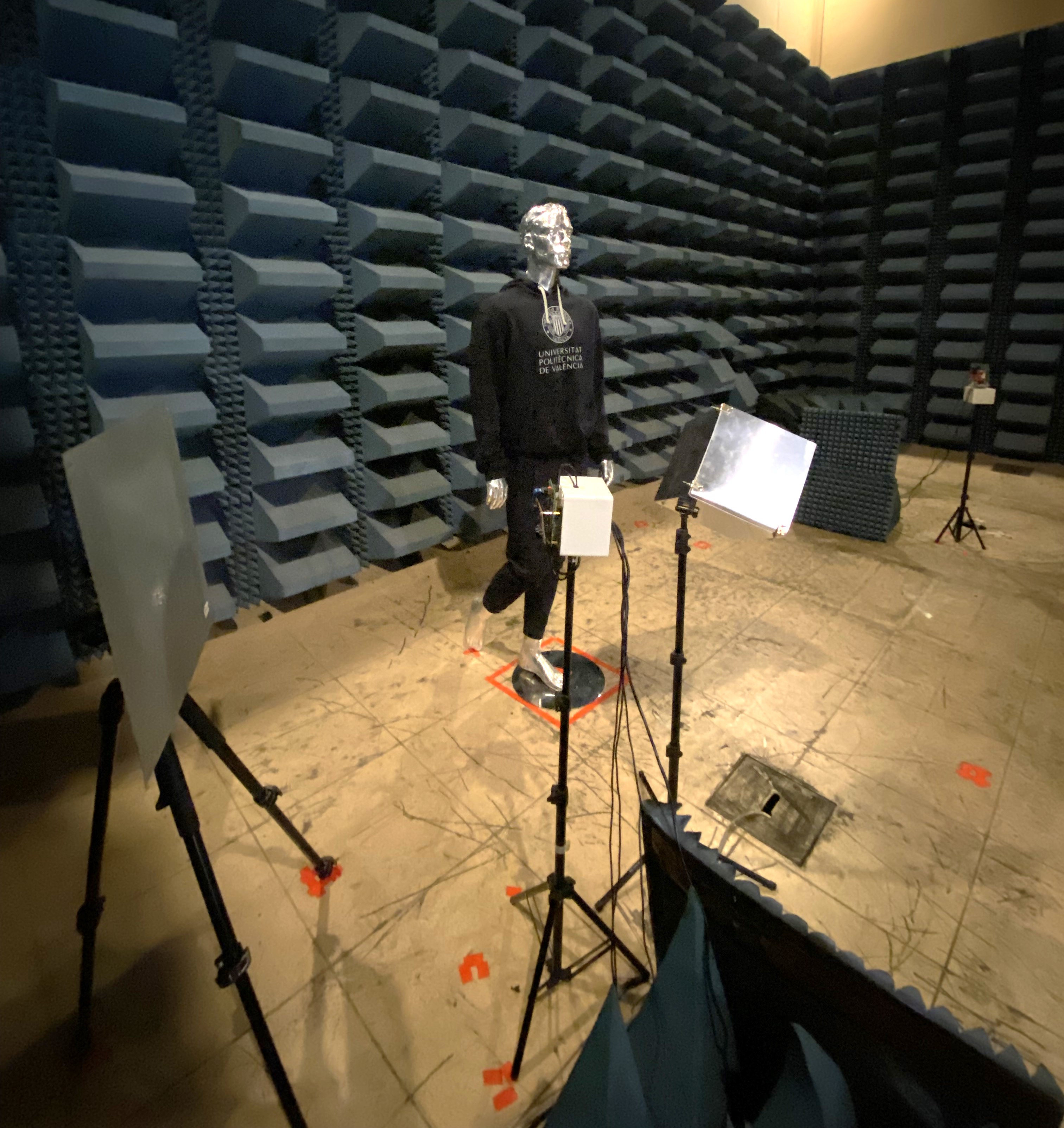}
  \caption{Set-up photography.}
  \label{fig:ph_S3}
\end{subfigure}
\caption{Measurement set-up 3.}
\label{fig:S3}
\end{figure}

The measurement campaign carried out for this paper consists of three controlled set-ups situated within an anechoic chamber. The proposed set-ups were selected to emulate a simple communication channel, which would facilitate geometrical analysis. Two objects were used as scatterers in the set-ups: a full-scale metalized mannequin (dummy) and a 50x50 cm metal plate held by a tripod. Moreover, the floor of the anechoic chamber was metallic, acting as an additional scatterer. In the subsequent sections, the conditions and geometry of each set-up are described.

In set-up 1, as depicted in Figure \ref{fig:S1_S2}, the metal plate and the dummy were placed to act as common scatterers. This resulted in several multipath contributions caused by the surfaces of both objects. 

In set-up 2, the metal plate was removed from the set-up, as shown in Figure \ref{fig:S1_S2}. Consequently, in the subsequent analysis, it can be observed how the multipath caused by the specular reflections on the metal plate disappeared.

Finally, a set-up to test the reliability of the techniques in Non-Line-Of-Sight (NLOS) situations is proposed. To generate multipath contributions in a controlled manner, a metal plate was placed in front of the receiving radar, creating an NLOS path from the transmitter reflected on the metal plate. Moreover, the floor creates multipath contributions and the dummy. Additionally, a corner reflector was positioned near the receiver to prevent multipath caused by secondary lobes of the radar antennas. This reflector was pointed to ensure that all the power was reflected back to the walls of the anechoic chamber. The schema for this set-up is shown in Figure \ref{fig:S3}.

\section{RESULTS AND ANALYSIS}
The tap-delay multipath components and their AoAs are used to verify the correct estimation of the communications channel. Additionally, RMS DS is obtained to select the area of interest in the signal.
\subsection{Measured RMS delay spread}
The RMS DS for the set-ups proposed earlier in \ref{set-up_desc} are obtained to be compared with the RT simulation described in section IV.b. The RMS DS is defined as: 
\begin{equation}
RMS\, DS = \sqrt{\frac{\sum_{i=0}^{N-1}{g_i\cdot(\tau _i-\bar{\tau})^2)} }{\sum_{i=0}^{N-1}{g_i}}}\label{eq:RMS DS}
\end{equation}

where $g_i$ corresponds to the amplitude of the $i^{th}$ sample and $\tau_i$ is the delay of the $i^{th}$ sample. The term $\bar{\tau}$ refers to the weighted average delays of channel components, which can be calculated as:
\begin{equation}
\bar{\tau} = {\frac{\sum_{i=0}^{N-1}{g_i\cdot\tau _i} }{\sum_{j=0}^{N-1}{g_j}}}\label{eq:averageDelay}
\end{equation}
RMS DS values are considered when comparing the channel with its RT simulation so that only the multipath components received within the delay spread time are considered for the comparison. 
\begin{table}[htbp]
\caption{RMS delay spread for the measured set-ups}\label{table:RMSDS}
\begin{center}
\begin{tabular}{|c|c|c|c|}
\hline
\textbf{RMS DS (ns)}&{\textbf{Set-up 1}}&{\textbf{Set-up 2}}&{\textbf{Set-up 3}} \\
\hline
\cline{2-4} 

\hline
\textbf{Rx 1}& 15.91 &10.56 & 36.45 \\
\hline

\hline
\textbf{Rx 2}&17.69 &14.09 &  37.89\\
\hline

\hline
\textbf{Rx 3}&12.70 & 6.67& 28.72 \\
\hline

\hline
\textbf{Rx 4}& 18.60 &15.98 & 37.13 \\
\hline

\end{tabular}
\label{tab1}
\end{center}
\end{table}

\begin{figure}
\centering
\begin{subfigure}{.5\textwidth}
  \centering
  \includegraphics[width=.9\linewidth]{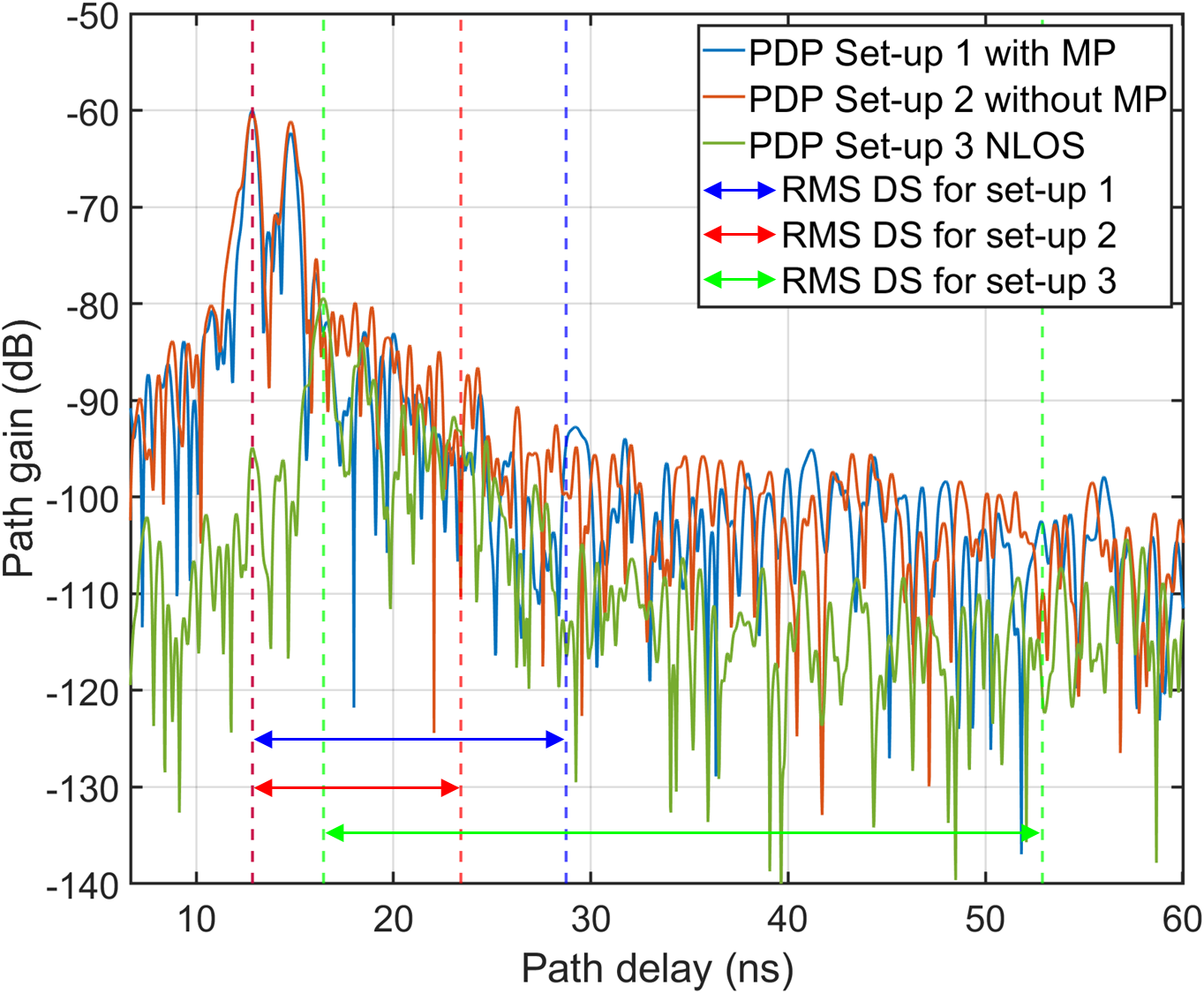}
\end{subfigure}%
\caption{RMS delay spread for the three set-ups.}
\label{fig:RMS_DS}
\end{figure}

Table \ref{table:RMSDS} contains data about the RMS DS values for each of the four receiving antennas and the three measured set-ups. The difference between the RMS DS of the four antennas is that the reflections are specular due to the type of scattering in this scenario, being very sensitive to the antenna's position. In the same manner, it is observed in the measurements in Figure \ref{fig:RMS_DS} that for the same scenario, the number of scatterers detected by every antenna can be different for the same reason previously mentioned. Therefore, performing a coherent integration of the signals from each receiver is essential, as described in Section \ref{algorithm}.

It can also be seen that, as expected, set-ups 1 and 2, both of LOS condition, have similar RMS DS values since the ground reflection overshadows the contributions from the metal plate. However, in set-up 3, which is an NLOS case, the RMS DS is much higher, as it corresponds to a rich multipath environment.  

\subsection{Comparison of the measured and Ray-tracing simulated channel}

The extracted channel is compared with an RT model to verify the channel-sounding technique presented in this paper. The RT model is based on Geometrical Optics (GO), forming the channel's simulated rays and a tap delay profile. The propagation equation calculates the received power of each tap (\ref{eq:trasmision_equation}), to which two loss terms are added, $L_{sys}$ referring to the losses caused by the system and $L_{ref}$ referring to the attenuation suffered by the reflection in each scatterer.

\begin{equation}
\frac{P_{rx}}{P_{tx}}=  \frac{G_{tx}\cdot G_{rx}}{ FSPL \cdot L_{sys} \cdot L_{ref}}
\label{eq:trasmision_equation}
\end{equation}
The rays considered in the simulation fall within the RMS DS obtained for each set-up. An alternative assumption for this comparison could be considering those taps that are within a certain margin of path gain (typically between 20 and 30 dB), but then some relevant components would not have been accounted for in some of the set-ups. 

\begin{figure}
\centering
\begin{subfigure}{.5\textwidth}
  \centering
  \includegraphics[width=0.95\linewidth]{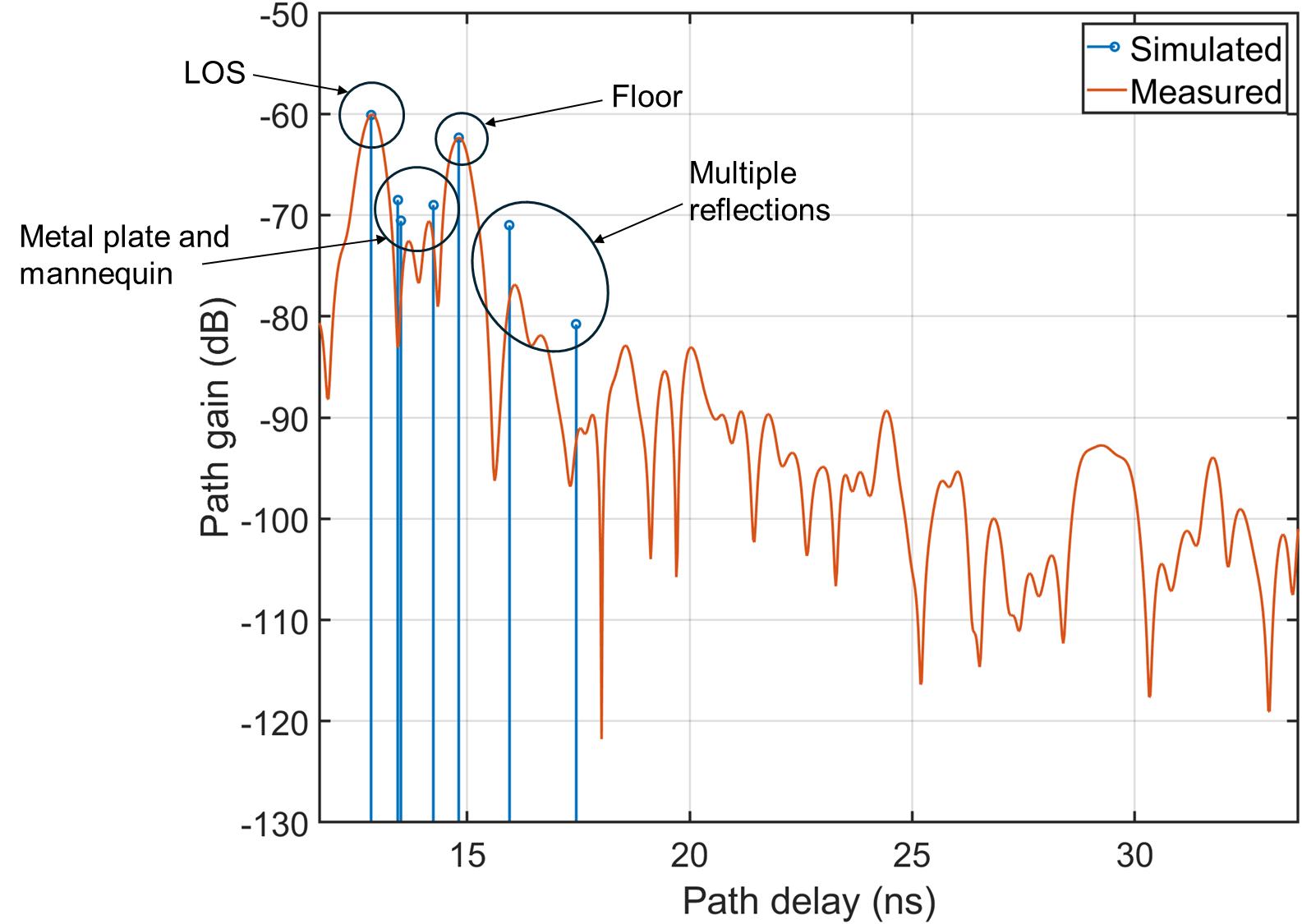}
\end{subfigure}%
\caption{PDPs of the set-up 1 measured and simulated channels.}
\label{fig:sim1}
\end{figure}

\begin{figure}
\centering
\begin{subfigure}{.5\textwidth}
  \centering
  \includegraphics[width=0.95\linewidth]{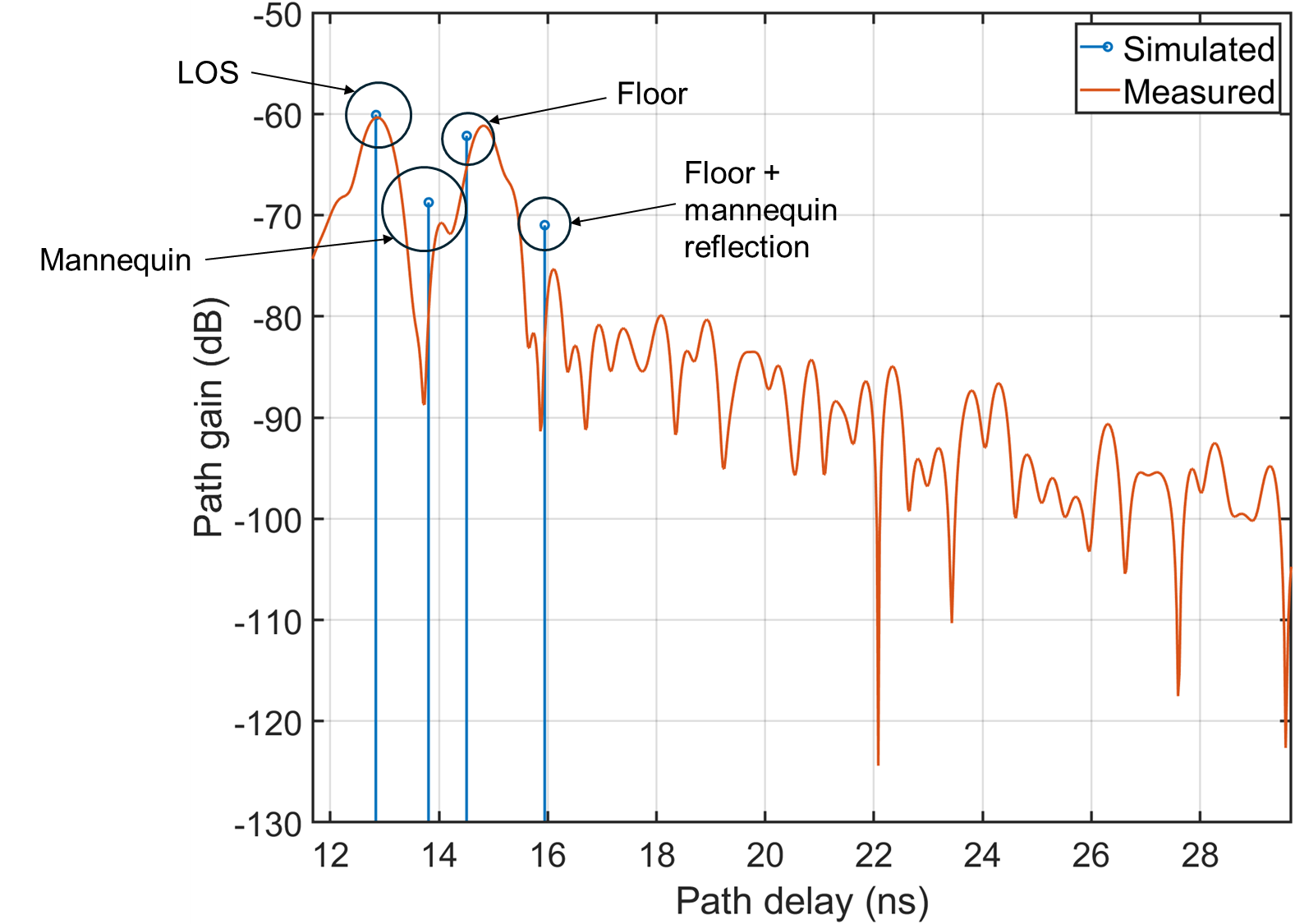}
\end{subfigure}%
\caption{PDPs of the set-up 2 measured and simulated channels.}
\label{fig:sim2}
\end{figure}

\begin{figure}
\centering
\begin{subfigure}{.5\textwidth}
  \centering
  \includegraphics[width=0.95\linewidth]{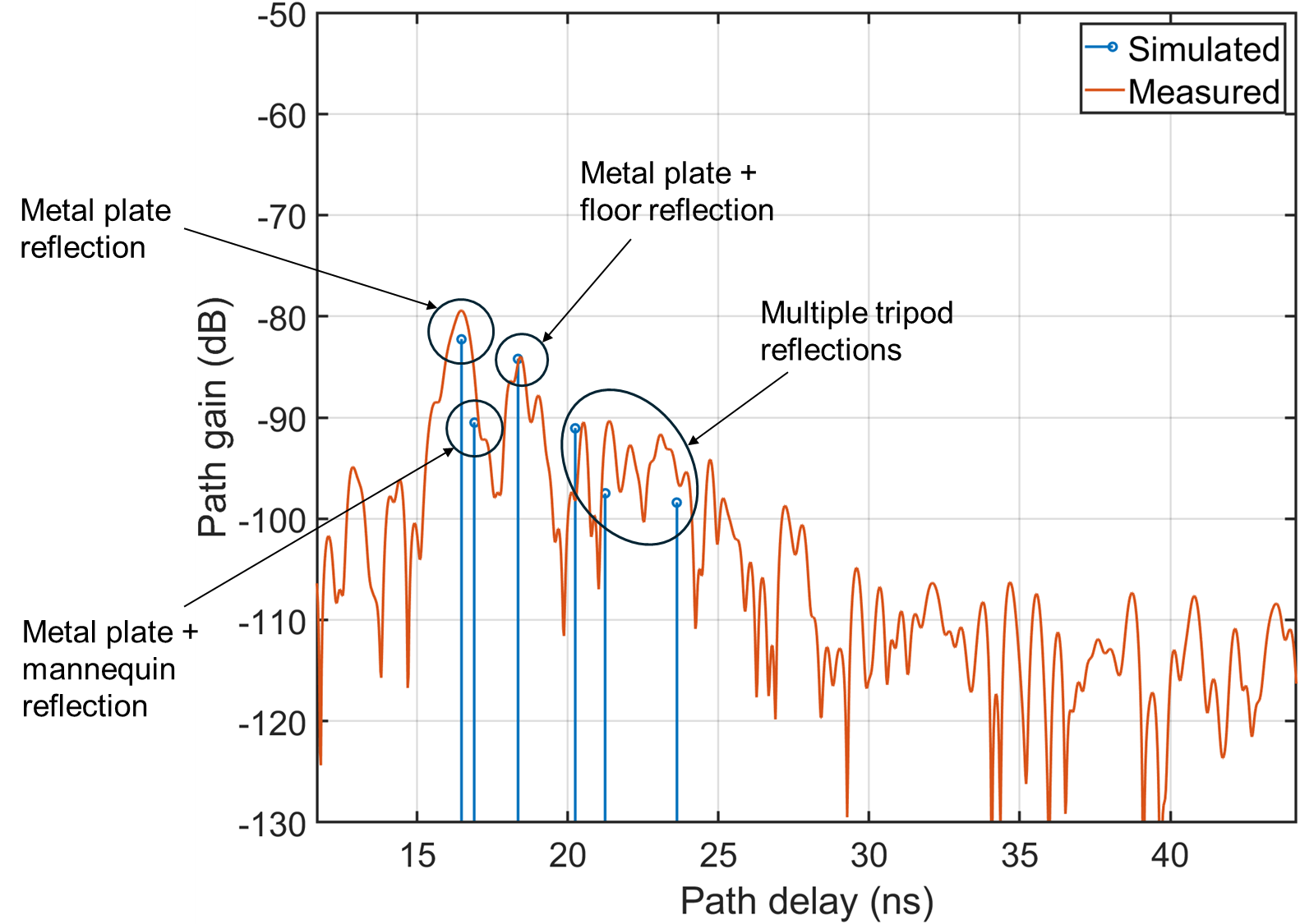}
\end{subfigure}%
\caption{PDPs of the set-up 3 measured and simulated channels.}
\label{fig:sim3}
\end{figure}

Figures \ref{fig:sim1}, \ref{fig:sim2}, and \ref{fig:sim3} compare the channels simulated for the three set-ups with the actual measurements. The LOS component was observed to play the most significant role in set-ups 1 and 2, followed by the floor reflection which acted as a specular surface. However, it was noticed that reflections from the metal plate and the dummy produced more attenuated reflections. Although the metal plate was a specular surface, it did not achieve perfect pointing, resulting in lower power for the multipath. On the other hand, the dummy, having a more irregular surface, produced several multipaths components caused by the reflection in different parts of the dummy. Moving on to set-up 3, the main multipath components originated from the reflection on the metal plate, and the other multipath taps came from multiple reflections on the metal plate and other scatterers. Interestingly, it was observed that the tripods of both the radars and some scatterers produced periodical reflections in this set-up.

\subsection{Angle of arrival analysis}

\begin{figure}
\centering
\begin{subfigure}{.5\textwidth}
  \centering
  \includegraphics[width=.72\linewidth]{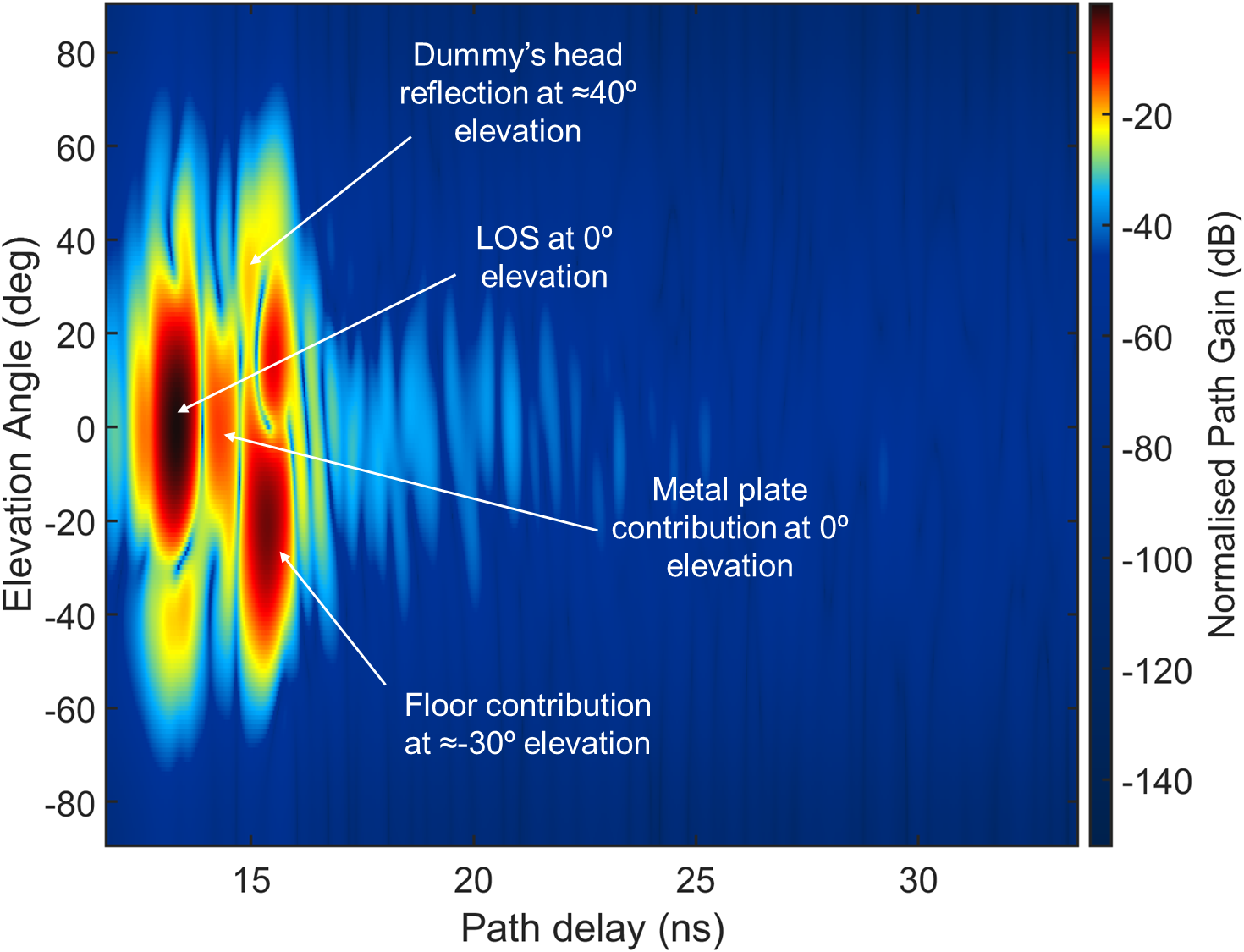}
  \caption{Elevation AoA for set-up 1.}
  \label{fig:AoA1}
\end{subfigure}
\newline
\vspace{2pt}
\begin{subfigure}{.5\textwidth}
  \centering
  \includegraphics[width=.72\linewidth]{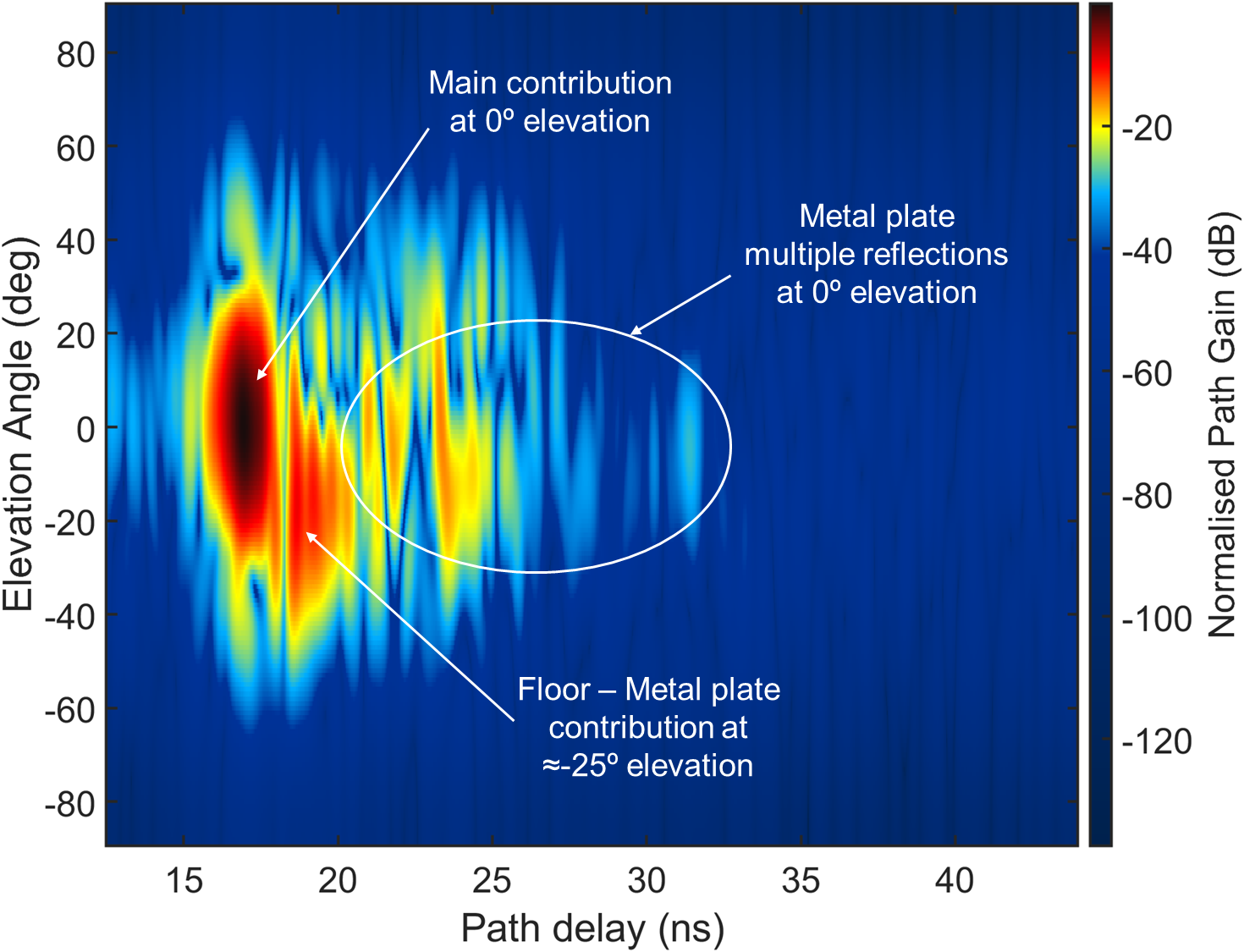}
  \caption{Elevation AoA for set-up 3.}
  \label{fig:AoA3}
\end{subfigure}

\caption{Angle of Arrival normalised PDPs distribution.}
\label{fig:AoA}
\end{figure}

As section \ref{RF_descr} mentions, the radar receiving antennas are positioned in a vertical array. This specific arrangement differentiates between the main components, which are the LOS and ground contributions, as they generate the primary multipath power. We can determine the components received for each angle and their respective distances by applying the angle-FFT. The results of the AoA analysis for each configuration are illustrated in Figure \ref{fig:AoA}.
The results of the angular distribution analysis demonstrate a correlation with the data obtained through simulation. The measured rays match their simulated origin. At an elevation angle of 0$^\circ$, the main multipath corresponding to the LOS can be observed in the case of set-ups 1 and 2. A strong reflection from the floor is also observed at -30$^\circ$, which matches the estimate obtained through RT. A smaller path can be observed at 40$^\circ$ between the LOS and the floor reflection contribution, which matches the reflection in the dummy's head. In the case of Figure \ref{fig:AoA3}, the main components arrive from reflections on the metal plate, while multiple reflections on the floor and metal plate can be observed at -25$^\circ$. Several additional multiple reflections are also verified to match with the simulated channels.

\section{CONCLUSIONS}
The paper aims to contribute to define dual-channel models for integrated sensing and communication (ISAC) and establishing their interrelationships. The study demonstrates the extraction of the bistatic sensing channel from monostatic measurements using radars in the mmWave band. Techniques for interference extraction, coherence analysis, module and phase correlation, chirp clustering, auto-clutter reduction, and ICZT are added to the previous works published by the authors. Measurements in the mmWave frequency band (77-81 GHz) for several set-ups that include LOS and NLOS conditions have been performed for comprehensive analysis. Comparison with RT simulations demonstrates the accuracy of the proposed channel extraction technique for the estimated PDP and AoA parameters. The proposed method for ISAC is effective in capturing communication channel properties accurately. The approach showcases promise for enhancing understanding and modeling of sensing and communication elements in emerging communication networks, particularly 6G.

\vspace{12pt}

\end{document}